\documentstyle[prb,aps,psfig,twocolumn]{revtex}

\input{psfig.sty}

\begin{document}

\draft

\title{Fermi-edge singularities in linear and nonlinear ultrafast spectroscopy}
\twocolumn[\hsize\textwidth\columnwidth\hsize\csname@twocolumnfalse\endcsname
\author{D. Porras, J. Fernandez-Rossier\cite{dagger} and C. Tejedor}
\address{Departamento de F\'{\i}sica Te\'orica de la Materia Condensada.
Universidad Aut\'onoma de Madrid. Cantoblanco, 28049 Madrid. Spain  }  
%\date{\today}
\maketitle

\begin{abstract}

We discuss Fermi-edge singularity effects on the linear and
nonlinear transient response of an electron gas in a doped
semiconductor. We use a bosonization scheme to describe the low
energy excitations, which allows to compute the time and temperature
dependence of the response functions.  Coherent control of the energy
absorption at resonance is analyzed in the linear regime. It is
shown that a phase-shift appears in the coherent control oscillations,
which is not present in the excitonic case. The
nonlinear response is calculated analytically and used to predict
that four wave-mixing experiments would present a Fermi-edge
singularity when the exciting energy is varied. A new dephasing
mechanism is predicted in doped samples that depends linearly on
temperature and is produced by the low-energy bosonic excitations in
the conduction band.     

\end{abstract}
]
\pacs{PACS numbers: 78.47.+p}

\narrowtext

\section{Introduction}

The promotion of an electron from a localized state in the valence
band to an empty state in a partially filled conduction band is
accompanied by a dynamical response of the Fermi gas. The enhancement
of the absorption probability when the new electron is promoted just
above the Fermi level is known as the  Fermi-edge singularity
(FES)\cite{Mahan}. This phenomenon has been observed in continuous
wave spectroscopy in a variety of doped semiconductor
heterostructures \cite{experiments,Herko}. FES arises as a result of
the interplay between two different physical processes: the sudden
appearance of a hole potential and the presence of an extra electron
at the conduction band. Both effects produce charge density
oscillations involving low energy electron-hole pairs. The
constructive interference between these two effects gives the FES.
Following the seminal work of Schotte and Schotte \cite{Schotte},
these low energy  electron-hole pairs can be described as Tomonaga
bosons. 

Coherent ultrafast spectroscopy of undoped semiconductors, where
excitons are the relevant excitation, has been much more widely addressed
than that of the doped case. Both linear and nonlinear techniques,
like Coherent Control (CC) and Four Wave-Mixing (FWM), have
been used to study the decay of the optical coherence induced by the
laser in undoped samples \cite{Shahbook}. In the case of doped
systems, only a few experiments has been performed. Kim {\em et
al.} \cite{Kim} carried out FWM experiments in n-doped GaAs
quamtum wells that presented FES in continuous wave spectroscopy.
In this experiment it was 
determined that  the carrier-carrier scattering rate was a
decreasing function of the exciting energy (above the Fermi energy),
in agreement with Landau theory. However,  the spectral width of
their laser pulses  was larger than the Fermi energy of the electron
gas so that Fermi edge excitations coexist with higher energy
electrons. Bar-Ad {\em et al.} performed FWM experiments  under
strong magnetic fields finding indications
of the nonlinear response of the FES. Brener {\em et al.}
\cite{Brener} performed  off-resonant pump and probe experiments in
n-doped GaAs QW, probing the ac Stark shift (a nonresonant
nonlinearity) in contrast with the works by Kim {\em et al.} and
Bar-ad {\em et al.} which measured resonant nonlinearities. From the
theory side, Perakis {\em et
al.}\cite{Perakis1,PerakisPRB,PerakisReport} have  studied the coherent
nonlinear response of the FES either under ultrashort laser pulses, or
under nonresonant excitation, {\em
i.e.}, when the nonlinearity comes from an intense laser pulse
spectrally peaked below the absorption threshold.  

Our work addresses a physical situation slightly  different from all
of the above: a doped semiconductor, in zero magnetic field, is
excited by  laser pulses spectrally peaked around the absorption
threshold, so that absorption takes place. Moreover, the laser pulses
are  spectrally narrow (compared to the Fermi energy $\epsilon_F$
measured from the bottom of the conduction band) so that the
photoexcited electrons have energies close to the Fermi level, but
the pulses are shorter than $T_2$  so that transient coherent effects can
be observed \cite{Shahbook}. 

 Our main
findings are: {\em i)} CC of the energy absorbed by the system  (the
analogous of CC of the exciton density \cite{CC,CC2,CC3}) can  be
performed in doped samples. CC oscillations show a characteristic
phase-shift which depends on the exponent of the continuous wave FES.
.  {\em ii)} The intensity of the FWM signal shows a singularity when the
exciting frequency is varied near the Fermi edge. {\em iii)} The
optical coherence induced by  the laser, both in the CC and FWM
situations,  has an intrinsic  {\em exponential} decay roughly
proportional to the temperature $T$. At zero $T$ the intrinsic decay
follows the well known power law associated to the FES in the linear
response\cite{Mahan}.

FES can be understood in a model of spinless free electrons which only
interact with a photoexcited hole \cite{Mahan,Nozieres,Ohtaka}. Within the 
Nozi\`eres-De Dominicis scheme
we consider a localized hole and a contact interaction:
\begin{eqnarray}
& H = & \sum_{k=0}^{k_D} \epsilon_k a^{\dagger}_k a_k 
+(E_g+\epsilon_F) \ d^ \dagger d +
\frac{V}{N} \sum_{k,k^{\prime}}^{k_D}
a_{k}^{\dagger}a_{k^{\prime}}d^{\dagger}d ,
\label{MND}
\end{eqnarray}
where ${d^{\dagger}}$ creates a localized hole 
and $\epsilon_k$ is the dispersion relation of electrons at the
conduction band, created by ${a_k^{\dagger}}$. 
${k_D}$ is a wave vector cut-off,  $V$ the attractive potential
between the hole and the electrons in the conduction band, and $N$
the linear size of the system. It must be stressed that two different
kind of excitations appear in the Hamiltonian (\ref{MND}): the
valence hole, and the conduction electron-hole pairs which can be
described, close to the Fermi energy, as bosonic excitations
\cite{Schotte}. These conduction electron-hole pairs are totally
unrelated to the excitons in undoped semiconductors,
 which involve both the conduction and the
valence band.

We discuss now some of the approximations involved in Hamiltonian
(\ref{MND}). First of all,  we assume that the valence hole has  an
infinite mass and it does not recoil in its interaction with the
conduction electrons.  Considering a finite mass hole would render
extremely difficult an analytical  calculation of the transient
nonlinear response.  In general, holes have a finite mass in real 
semiconductors. However, there is a number of situations in which the
hole can behave as an infinite mass particle. Strong localization of
the holes can happen  due to both alloy fluctuations in general and 
single monolayer fluctuations in narrow quantum wells. The hole is
also strongly localized in the case of  'acceptor to conduction band'
transitions  in an n-doped  semiconductor slightly compensated with
acceptor impurities like Beryllium \cite{Herko}. From the theory
point of view, it is well established that the finite mass of the
valence hole  reduces the FES,  especially in emission
\cite{Uenoyama}.  Hence, the  experimental observation of FES in a
real system supports  the existence of strongly localized valence
holes. 

Second, Hamiltonian (\ref{MND}) only includes a single valence hole.
This is known to give the correct linear response in semiconductor
samples. Nevertheless, in the case of excitons, two valence hole
states must be included in order to get the correct third-order
optical response \cite{chi3excitons}, which has a contribution coming
from  the exciton-exciton
interaction. Note, however, that (\ref{MND}) is analogous to a
two-level system (the valence hole) dressed by the final-state
interaction with the Fermi sea electrons. This implies that the
single valence hole case presents optical nonlinearities which do
not exist in the excitonic case, and govern the nonlinear response in
a low excitation regime.  The most important process which
invalidates this approximation is the overlap between the different
perturbations induced on the conduction electrons by valence holes at
different sites. The range of this perturbation can
be estimated as $k_F^{-1}$. The overlap will be negligible if the
density of photoexcited valence holes ($n_{vh}$) is low enough, so
that  the distance between valence holes is greater than $k_F^{-1}$. 
This is the case for typical excitation densities of $10^9$, $10^{10}
cm^{-2}$ in FWM experiments in doped GaAs quantum wells with a Fermi
energy of 20 meV, so that $k_F^{-1} n_{vh}^{1/2} \approx 10^{-2}$. In
this range the Coulomb interaction between carriers at different
valence hole sites can also be neglected. Under these conditions, the
optical response of a sample with many valence holes will be
equivalent to the optical response of Hamiltonian (\ref{MND}).

This paper is organized as follows: In section II we review the bosonization
approach to the FES linear response. Our original contribution starts in
subsection II.C, where we use this approach to obtain the nonlinear optical
response of the FES. In section III we discuss the predictions of the linear
response theory at finite temperature in the case of a CC experiment. In
section IV we apply our calculation of $\chi^{(3)}$ to the case of
various FWM experiments. The discussion of our results is made in section V,
where we consider the comparison of the dephasing mechanisms contained in
(\ref{MND}) with other competing processes.

\section{theoretical framework}

\subsection{The bosonization scheme}

Since only states close to the Fermi
level are excited, we can approximate the dispersion relation by 
${\epsilon_k = (k-k_F)/\rho}$
with ${\rho}$ being the density of single particle states at the Fermi level. 
We consider a hole potential $V$ isotropic and weak,
so that only s-wave scattering is important. Under these conditions, 
the problem becomes that of one dimensional
electrons with linear energy dispersion. 
The bosonization approach allows to express all the 
physics in terms of the bosonic fields,
\begin{equation}
b^{\dagger}_k = \sum_{k^{\prime}=k}^{k_D} 
a^{\dagger}_{k^{\prime}}a_{k^{\prime}-k}/ \sqrt{kN} .
\label{bozoniz1}
\end{equation} 
with  $0 \leq k \leq k_F$. 
The set of operators
${b_k}$, ${b^{\dagger}_k}$ satisfies bosonic commutation relations only when
one restricts to the low-energy range\cite{Mahan}. We define $H_i$ as the
initial Hamiltonian without a valence hole ($d^{\dagger}d=0$) and 
$H_f$ as the final Hamiltonian after the photoexcitation of the valence
hole ($d^{\dagger}d=1$). They can be written in terms of the bosonic operators
\cite{Schotte}: 
\begin{eqnarray}
& H_i = & \sum_k \frac{k}{\rho} b^{\dagger}_k b_k
\nonumber \\ 
& H_f = & \omega _0 + \sum_k \frac{k}{\rho} 
(b^{\dagger}_k+\frac{\rho V}{\sqrt{kN}})(b_k+\frac{\rho V}{\sqrt{kN}}) .
\label{bosoniz2}
\end{eqnarray}
where $\omega_0=E_g+\epsilon_F-(V\rho)^2\epsilon_F$ is the renormalized 
hole energy (we set $\hbar = 1$). The index $k$ in $b^{\dagger}_k$, $b_k$ always
runs between $0$ and $k_F$.

$H_i$ and $H_f$ are related by a canonical transformation which describes the
effect of the potential created by the valence hole onto the conduction 
electrons:

\begin{equation}
H_f = \omega_0 + U^{\dagger} H_i U .
\label{transformation}
\end{equation}

where,

\begin{equation}
U = exp \left[ V \rho \sum_k \frac{1}{\sqrt{kN}} (b^\dagger_k-b_k)
 \right] .
\label{canonical}
\end{equation}

Optical properties are determined from the adequate correlation functions
of the electric dipole operator $P^{\dagger} = \mu \ a^{\dagger}
d^{\dagger}$ , where ${\mu}$ is the dipole matrix element and
$a^{\dagger}$ is the creation operator of conduction electrons at the localized
hole site. This operator can also be expressed as an exponential of Tomonaga
boson operators: 

\begin{equation}
a^{\dagger}=\sum_{k=0}^{k_D}a_k^{\dagger} 
= exp\left[ \sum_k \frac{1}{\sqrt{kN}}(b^{\dagger}_{k}-b_k) \right] .
\label{bosoniz3}
\end{equation}

\subsection{Linear response}

The linear response $\chi^{(1)}(t)$ is given (in the Rotating Wave Approximation)
by the expression:

\begin{equation}
\chi^{(1)} (t) = i \mu^2 \theta (t) \langle P(t) P^{\dagger} (0) \rangle .
\label{kubo}
\end{equation} 

$P^{\dagger}$ creates a valence hole, so that the system evolves under the final
Hamiltonian $H_f$ in the interval $(0,t)$:

\begin{eqnarray}
\langle P(t) P^{\dagger} (0) \rangle & = & 
 \langle e^{i H t} P(0) e^{- i H t} P^{\dagger} (0)  \rangle = \nonumber \\
 & & \langle e^{i H_i t} a U^{\dagger} e^{- i H_f t} U a^{\dagger}  \rangle 
  e^{-i \omega_0 t} = \nonumber \\
 & & \langle B(t) B^{\dagger} (0) \rangle
  e^{-i \omega_0 t} ,
\label{evolution}
\end{eqnarray} 

where

\begin{equation}
B^{\dagger}(t) = 
exp[(1+ V \rho) \sum_k \frac{1}{\sqrt{kN}}
 (b^{\dagger}_k e^{i \frac{k}{\rho} t} - b_k e^{- i \frac{k}{\rho} t})] .
\label{Bes}
\end{equation}

The original Schotte and Schotte result \cite{Schotte} can be extended to the case
of nonzero temperature by considering a bath of Tomonaga bosons at thermal
equilibrium in the average (\ref{evolution}):

\begin{eqnarray}
 \langle B(t)B^{\dagger} (0) \rangle &&= \nonumber \\ 
   exp [ - && \sum_{k} \frac{(1+V \rho)^2}{kN} \times  \nonumber \\
     ( (1&&+2 N_B(k))
   (1-Cos(\frac{k}{\rho}t))  + i Sin(\frac{k}{\rho}t)] ) ]  ,
\label{average}
\end{eqnarray}

where $N_B(k)$ is the Bose-Einstein occupation factor. We are interested in the
long-time limit of the response functions. The cut-off in momentum space in
(\ref{average}) is $k_F$ and it implies a cut-off in energy space,
$\epsilon_c = k_F/\rho = 2 \epsilon_F$, as usually taken in the bosonization
procedure. In the limit $t>>\epsilon_c^{-1}$ we obtain:

\begin{equation}
\chi^{(1)}(t)=i \mu^2 \theta(t)  
\left[i \epsilon_c \frac{Sinh(\pi k_B Tt)}{\pi k_B T} \right]^{-\alpha} 
e^{-i \omega_0 t} , 
\label{chi1}
\end{equation}

where ${\alpha}=(1+V \rho)^2$.  Expression (\ref{chi1}) will be valid
in the case of near-resonance excitation and spectrally narrow
pulses, that is, $| \omega - \omega_0|<<\epsilon_c$, and $(\Delta
t)^{-1}<<\epsilon_c$  ($\omega$ is the excitation energy). Condition
$k_B T << \epsilon_c$ must also be fulfilled in order to consider low
energy excitations only.  At zero $T$, Eq. (\ref{chi1}) recovers the
well known behavior $\chi^{(1)}(t)= i \mu^2 \theta(t) (i \epsilon_c
t)^{-\alpha}$. 

In the spectral domain, the absorption is given by
$\theta(\omega-\omega_0) (\omega-\omega_0)^{(\alpha-1)}$ so that FES
takes place for $\alpha<1$.  In the time domain, the FES is 
characterized by the {\em intrinsic} power law decay of the response
function (with $\alpha <1$). As we show below, the decay of the
optical coherence, {\em i. e.}, the dephasing, is an increasing
function of  $\alpha$, which is the square of a sum of two  terms
which have different physical origin and opposite effects. The first
term,  $1$, is related to the addition  of a new electron to  the
Fermi level in the absorption  process. The second term, $-|V \rho|$,
is related to the sudden switching of the hole potential.  The first
term makes dephasing {\em more} efficient while the second one makes
dephasing {\em less} efficient.

\subsection{Third-order susceptibility}

To study the nonlinear response of the electron gas we concentrate 
in  FWM experiments, which are usually described by means of the third-order
susceptibility, $\chi^{(3)}$. However it is not evident whether a perturbative
expansion in
terms of the electric field is justified in the case of the nonlinear optical
response of Hamiltonian (\ref{MND}). Primozich {\em et al.} 
\cite{PerakisPRB,PerakisReport}
have shown the validity of such an expansion provided that 
$(\mu {\mathcal{E}}_0 \Delta t)^2<<1$.
Considering excitation
intensities of mW, 
$\Delta t = 0.7$ps and known values for the interband dipole
matrix element of GaAs \cite{Bastard}, one obtains
$(\mu {\mathcal{E}}_0 \Delta t)^2 \approx 10^{-3}$. Thus, we can consider terms up
to the third order in the electric field for the ultrafast transient experiments
described below.

We
consider the typical situation in which the system is excited by two mutually
delayed laser pulses which propagate along
different directions, ${\bf k_1}$ and ${\bf k_2}$, with $|{\bf k_1}|=|{\bf
k_2}|$. In any system with translational invariance and some degree of 
nonlinearity in the optical response, these exciting pulses will induce an
electric dipole which will re-emit light along the direction $2 {\bf  k_2}
-{\bf k_1}$. Up to the third order in the external field the FWM signal 
is given by:
\begin{eqnarray}
& F_{FWM}(t)= & \int _{- \infty} ^t \!\!\! dt_1 dt_2 dt_3      
\chi^{(3)} (t-t_1,t-t_2,t-t_3)   
\nonumber \\
& & \times E^{\ast}_1(t_2)E_2(t_1)E_2(t_3) + h.c. ,
\label{nlsignal1}
\end{eqnarray}
where $E_{1,2}$ are the electric fields in the directions $\bf k_{1,2}$.
As in the case of CC, FWM takes place as long as the polarization induced by the
first laser pulse is not wiped out before the second pulse reaches the sample.
For this reason, both CC and FWM can be used to measure $T_2$. In 
undoped samples $\chi^{(3)}$ is related to the exciton-exciton interaction. In
the case of the FES we are going to see that $\chi^{(3)}$ is not zero even for
non-interacting electrons. This constitutes an important difference between the 
doped and undoped systems.

Performing a 
perturbation expansion up to third order in the electric field, it can be shown
that 
$\chi^{(3)}$ is proportional to the average of four polarization operators
\cite{Mukamel}:

\begin{eqnarray}
 \chi^{(3)} (t-t_1,t-t_2, && t-t_3) = -i \nonumber \\
\left( \theta(t-t_1) \right. && \theta(t_1-t_2) \theta(t_2-t_3)  \nonumber \\
 && \langle P(t) P^{\dagger}(t_1) P(t_2) P^{\dagger}(t_3) \rangle  \nonumber \\
 +
       \theta(t-t_1) && \theta(t_1-t_2) \theta(t-t_3) \nonumber \\
 && \langle P(t_2) P^{\dagger}(t_1) P(t) \left. P^{\dagger}(t_3) \rangle \right) .
\label{chi3general}
\end{eqnarray}

In $\langle P(t) P^{\dagger}(t_1) P(t_2) P^{\dagger}(t_3) \rangle $
the second and fourth polarization operators create a valence
hole, so that the system evolves under $H_f$ inside the intervals $(t_2,t_3)$ and
$(t,t_1)$. Using the same argument that
leads to Eq. (\ref{evolution}) it is straightforward to show that

\begin{eqnarray}
\langle P(t) && P^{\dagger}(t_1) P(t_2) P^{\dagger}(t_3) \rangle \nonumber \\ 
 = && \mu^4 \langle B(t) B^{\dagger}(t_1) B(t_2) B^{\dagger}(t_3) \rangle
 e^{-i (t-t_1+t_2-t_3) \omega_0 } .
\label{fourp}
\end{eqnarray}

Using the definition of $B^{\dagger}$ given by Eq. (\ref{Bes}) we can
express $\chi^{(3)}$ as the thermal average of a product of four exponentials of
bosons. The average of a product of any number of 
exponentials of bosons can be factorized
into two-exponential correlation functions. In Appendix A this fact is used to
prove the general result:

\begin{eqnarray}
\langle B(t_0) B^{\dagger} (t_1) &&... B(t_{m-1}) B^{\dagger}(t_m) \rangle
\nonumber \\ = &&
\prod_{j>i=0}^n \langle B(t_i) B^{\dagger}(t_j) \rangle ^{(-1)^{1+i+j}} ,
\label{chin}
\end{eqnarray}

where n is an odd integer.
$\langle B(t_i) B^{\dagger}(t_j) \rangle$ is given by (\ref{average}) and,
in the long time approximation, by (\ref{chi1}). 
The nonlinear susceptibility $\chi^{(n)}$ at
any order $n$ can be expressed by means of products of $n+1$ polarization
operators of the form $\langle P P^{\dagger}...P P^{\dagger}\rangle$.
Each polarization operator can be expressed as an exponential of bosonic
operators.
Thus, Eq. (\ref{chin}) allows the
calculation of the optical response of the FES 
at any order in the electric field in the long-time limit, under the
approximations discussed in section I.

Application of (\ref{chin}) to the case of $\chi^{(3)}$ yields the result:

\begin{eqnarray}
 \langle && P(t)P^{\dagger}(t_1)P(t_2)P^{\dagger}(t_3) \rangle = \mu^4 \times
\nonumber \\ 
 && \frac{\langle B(t)B^{\dagger}(t_1) \rangle 
\langle B(t_1)B^{\dagger}(t_2) \rangle 
\langle B(t_2)B^{\dagger}(t_3) \rangle 
\langle B(t)B^{\dagger}(t_3) \rangle }
{\langle B(t)B^{\dagger}(t_2) \rangle 
\langle B(t_1)B^{\dagger}(t_3) \rangle } \times \nonumber \\
&& \hspace{5cm}  e^{-i (t - t_1 + t_2 - t_3) \omega_0} .
\label{polariz}
\end{eqnarray} 

This result implies that $\chi^{(3)}$ will present  
singularities similar to that of $\chi ^{(1)}$. Using
the result of equation  (\ref{chi1}) in equation 
(\ref{polariz}) we obtain the following $T=0$
expression for  $\chi^{(3)}$:
\begin{eqnarray}
\chi^{(3)} \propto \left[ \frac{(t-t_1)(t_1-t_2)(t_2-t_3) 
(t-t_3)}{(t-t_2)(t_1-t_3)} \right] ^{-\alpha} \, . 
\label{powers}
\end{eqnarray}

This simple expression is valid for $0 \leq \alpha <1$. Out of this
range the expression is more complicated. 
Equations (\ref{nlsignal1}), (\ref{chi3general}) and (\ref{polariz}) allow us to
calculate the FWM signal in near-resonance experiments, under the same conditions
explained under Eq. (\ref{chi1}).

\section{Linear Response: Coherent Control experiments}

In CC experiments, the sample is excited by a pair of phase locked identical laser
pulses delayed in a time $\tau$ with respect to each other. The total energy
absorbed by the system, $W$, as a function of the delay $\tau$
can be measured by detecting the reflectivity changes produced by the
photoexcitation density \cite{CC} or by measuring the total luminiscense
emitted
by the sample \cite{CC2}. These experiments are carried out in the linear
regime, where the total energy absorbed after photoexcitation 
can be easily calculated by means of the
linear response function:

\begin{equation}
W(\tau) = 2 Im \int_{- \infty}^{\infty}
 \chi^{(1)}(t_1 - t_2) E^{*}(t_1)E(t_2) dt_1 dt_2 . 
\label{CC}
\end{equation}

The  electric field of the phased locked laser pulses is given by 
$E(t)= {\mathcal{E}}(t)e^{-i \omega_0 t} +{\mathcal{E}} (t-\tau)e^{-i
\omega_0 (t-\tau)}.$ The pulses are thus spectrally peaked around 
the FES transition. The envelope functions are Gaussian pulses of
width $\Delta t$: ${\mathcal{E}}(t) = {\mathcal{E}}_0 e^{-t^2/\Delta
t^2}$. Substituting the electric field into the expression (\ref{CC})
it can be clearly seen that $W(\tau)$ depends strongly on $\tau$. It
oscillates with frequency $\omega_0$, showing that  the absorption
in  doped semiconductors,  close to a Fermi-edge singularity,  can be
coherently controlled. The phase and the amplitude of these
oscillations change also with $\tau$. We can distinguish three
different regimes:

{\em{(1)}} For $\tau < \Delta t$, the two pulses overlap:
the absorbed energy oscillates between $0$ (destructive interference)
and $4 W_{SP}$ (constructive interference),

\begin{equation}
W(\tau)= 2 W_{SP} (1 + cos(\omega_0 \tau)) ,
\label{shorttau}
\end{equation}

where $W_{SP}$ is the
energy transfered by a single pulse.

{\em{(2)}} For $\tau >> \Delta t, 1/\pi k_B T$, the decay of the
polarization between the two pulses is exponential, as can be seen
clearly in the behavior of $\chi^{(1)}$ for long times. It can be
easily proved  that in this regime,

\begin{equation}
W(\tau) = 2 W_{SP} + 
     W_{CC} e^{-\alpha \pi k _B T \tau} cos(\omega_0 \tau + 
      \alpha \frac{\pi}{2})
     ,
\label{longtau}
\end{equation} 

where $W_{CC}$ is the constant prefactor before the exponential decay and is given
by:

\begin{equation}
W_{CC}= 2 \pi (\Delta t)^2 \left(\frac{\epsilon_c}{2 T} \right)^{-\alpha}
exp(\frac{1}{2} (\alpha \Delta t k_B T)^2) .
\label{wcc}
\end{equation}

In the general case $W_{CC} \neq 2 W_{SP}$, 
due to the finite width of the exciting pulses and the fact that the decay is
non-exponential for short times. Eq. (\ref{longtau}) shows important
differences with the case of CC of excitons. First of all, a
phase-shift of $\alpha \, \pi/2$ appears in the CC oscillations at
long $\tau$. This surprising behavior is not observed in undoped
samples \cite{CC}, where the maxima of the oscillations are exactly
at  $\tau=2 n \pi /\omega_0$. The great interest of this phase-shift
in the CC oscillations resides in the fact that it is independent
on the relative importance of other competing dephasing processes.
This could allow a more accurate determination of the FES  exponent,
$\alpha$, than  in continuous wave photoluminescence
experiments.

Second, the exponent of
the  coherence decay behaves linearly with temperature, with the
factor $\pi \alpha$. In section V it is shown that this one is the
most  important temperature dependent dephasing mechanism at low
temperatures. Thus, the measure of the decay time of the CC
oscillations could allow another independent determination of the
singularity exponent. 

{\em{(3)}}
For very low temperatures, we can have $ 1/\pi k_B T >> \tau >> \Delta t$.
In this range, the decay of the polarization is non-exponential, even when the
pulses do not overlap, because of the
behavior of $\chi^{(1)}$ at short times. However, the condition 
$\tau,\Delta t> {\epsilon_c}^{-1} $ can
still be fulfilled, so the asymptotic approximation that leads to Eq. (\ref{chi1})
is valid. This non-exponential decay is another important difference with the undoped
case.

\begin{figure}
\psfig{figure=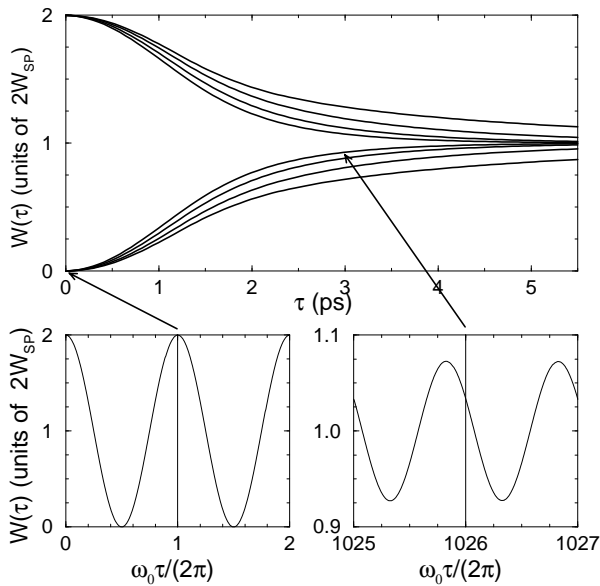,height=8.0cm,width=8.0cm}
\caption{Upper panel: evolution of the absorbed energy as a function of $\tau$ for
Gaussian pulses of width $\Delta t=0.7$ps, and $\alpha = 0.7$, $\omega_0=1.5$eV. 
Only the envelopes of
the CC oscillations are plotted, corresponding to temperatures between 1 K (outer)
and 4 K (inner). Lower panels: CC oscillations at $\tau=0$ps (left) and $\tau=3$ps
(right) for the case $T=4$K. In the right panel the maxima of the oscillations are
not at integer values of $\omega_0 \tau /(2 \pi)$, showing a characteristic
phase-shift.}
\label{cc}
\end{figure}

In an intermediate region of parameters,
the integration in (\ref{CC}) must be performed numerically.
The result of this calculation is presented in Fig. \ref{cc},
for $T=1-4K$, and clearly shows
the different regimes and the phase-shift $\alpha \, \pi/2$ and exponential
relaxation for long $\tau$.

The main conclusion from this section is that the absorption  in
doped semiconductors, close to the FES, can be coherently controlled.
The decay of the polarization predicted in cases {\em{(2)}} and
{\em{(3)}} is not produced by any inelastic scattering mechanism or
some sort of inhomogeneous broadening as it happens in the CC of
excitons \cite{CC3}. Instead, it is an {\em intrinsic} effect due to
the excitation of a continuum of bosonic   modes with a distribution
of energies which implies destructive interference in the time
domain.  In the absence of the potential created by the photoexcited
hole ($V=0$), this effect has been described as inhomogeneous
broadening in momentum space \cite{Kim}. However, the sudden
switching of the hole potential  partially compensates the effect of the
momentum space broadening, reducing the dephasing. 
This situation resembles that of the experiment of Wehner {\em et al.}
\cite{Wehner}, where the electron-LO phonon scattering rate is
coherently controlled. In our case, the Tomonaga bosons play the role
of the phonons in that experiment, with an important difference: the
Tomanaga bosons form a gapless continuum of
modes, which leads to the dephasing of the optical polarization.

\section{Nonlinear response: FWM experiments}

In this section we study the usual transient FWM experiments in which
the exciting fields appearing in Eq. (\ref{nlsignal1}) are
${E_{1,2}(t)={\mathcal{E}}_{1,2} (t) e^{-i\omega t}}$, where
${\mathcal{E}}_{1,2} (t)$ are Gaussian pulses of width $\Delta t$,
delayed in $\tau$ with respect to each other 
(${\mathcal{E}}_2(t-\tau)={\mathcal{E}}_1(t)$), and $\omega$ is the
central exciting frequency, which is taken at the FES resonance. 

\subsection{Decay of the FWM intensity with temperature}

We are now interested in the properties of the nonlinear optical
response, rather than in the dephasing processes between the pulses.
Therefore, in subsection IV.A and B, we take $\tau=0$.  From the
factorization formula for $\chi^{(3)}$ given in Eq. (\ref{polariz}),
we expect to find, in a FWM experiment, some of the characteristics
of the FES in linear response, such as a strong dependence on
temperature. 

\begin{figure}
\psfig{figure=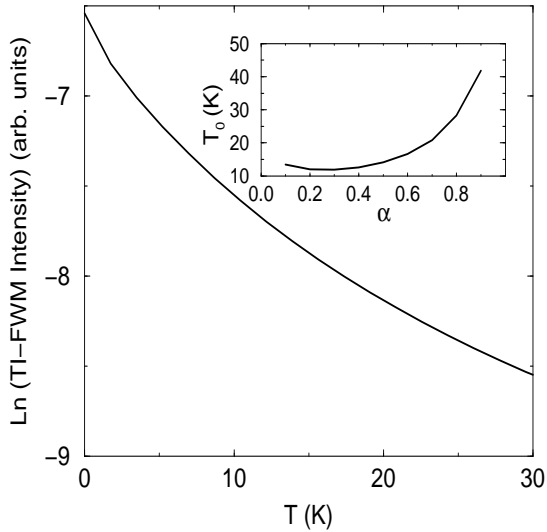,height=8.0cm,width=8cm}
\caption{Temperature dependence of the TI-FWM signal for $\alpha =0.7$,
$\tau =0$ and $\Delta t=0.7$ps. 
The inset shows the exponential decay parameter $T_0$ between $10$ and $30$K
as a function of $\alpha $.}
\label{temperature}
\end{figure}

In order to test this idea we calculate the time-integrated FWM
(TI-FWM) intensity, $I_{FWM}=\int dt |F_{FWM}(t)|^2$ when the sample
is excited at resonance ($\omega = \omega_0$) by Gaussian pulses with
$\Delta t=0.7$ps. In Fig. \ref{temperature}  we present our results
for the particular case $\alpha=0.7$, as a function of temperature.
We focus on the interval between 10 and 30 K, for comparison with
experiments \cite{Bar} (at higher $T$ the condition $k_B T <<
\epsilon_c$ is not satisfied). In this range our result for the decay
with temperature can be fitted to an exponential form $
e^{-T/T_0(\alpha)}$, so that we can obtain a characteristic
temperature $T_0(\alpha)$ which governs the decay of the FWM signal.
The parameter $T_0$ is as a function of $\alpha$  in
the inset of Fig. \ref{temperature}. A similar exponential decay of
the TI-FWM of a doped sample under high magnetic field has been
observed by Bar-Ad {\em et al.} \cite{Bar}. If we apply our zero
magnetic field theory to their result, we would infer $\alpha \approx
0.7$, a good value to get FES as the ones observed in continuous wave
spectroscopy \cite{experiments}. This could be a hint that the
physics of the FES under magnetic fields could be described by a
model similar to the one presented here, but further work is needed
to clarify this point.  

\subsection{FWM intensity as a function of the exciting frequency}

\begin{figure}
\psfig{figure=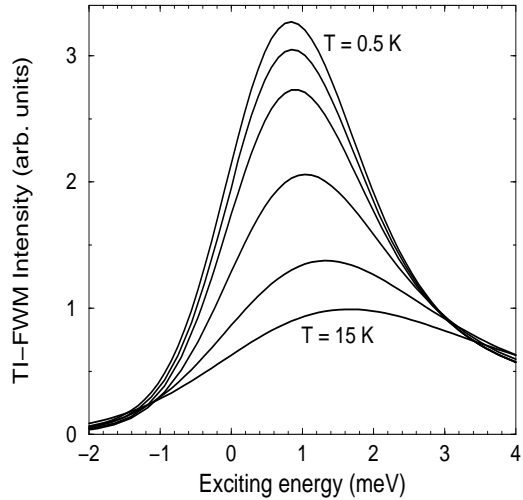,height=3.in,width=3.in}
\caption{TI-FWM signal as a function of the exciting frequency $\omega-\omega_0$,
for the case $\Delta t = 0.7$ps, $\alpha=0.7$, and different temperatures $T = 0.5,
1, 2, 5, 10, 15$ K (from top to bottom).}
\label{nonres}
\end{figure}

Now we treat the case in which $\tau=0$ and the exciting pulses are
slightly out of resonance ($\omega \neq \omega_0 $). Condition
$|\omega-\omega_0|<<\epsilon_c$ must be satisfied in order for the
bosonization procedure to be valid. We have calculated the TI-FWM
intensity as a function of $\omega$ for $\Delta t = 0.7$ps, $\alpha
=0.7$, and different temperatures. Our results are presented in Fig. 
\ref{nonres}, where it is shown that the FES appears as an
asymmetric resonance in the FWM spectrum, similar to the one which is
observed in linear spectroscopy. The FES resonance is strongly
suppressed with temperature and shows Lorentzian broadening for high
$T$, as expected from the exponential decay of $\chi^{(3)}$ at long
times. This fact allows to unambiguously determine the observation of
the FES in the nonlinear regime. A strong resonance in the FWM signal
as a function of the exciting frequency was reported in the work of
Kim {\em et al.} \cite{Kim}, in a doped sample which also showed a
FES resonance in the photoluminescence experiments.   

\subsection{FWM signal as a function of $\tau$}

In a transient degenerate FWM experiment the nonlinear signal can be
studied as a function of the time delay, $\tau$. We consider first
the time resolved  FWM (TR-FWM) signal, $F_{FWM}(\tau,t)$, which is
 a function of both the detection time, $t$ and the delay $\tau$.
  $F_{FWM}(\tau,t)$ can be
estimated  by assuming that the laser pulses amplitudes can be
approximated by $\delta$-functions (obviously, this is justified in
the case $t>>\tau>>\Delta t$). Using this assumption, simple
analytical expressions can be obtained:
\begin{eqnarray}
 &F&_{FWM} (\tau, t) \propto -i\mu^4(i\epsilon_c)^{-3 \alpha} \theta (\tau) 
\theta (t- \tau ) e^{-i \omega _0 t}
\nonumber \\
&& \times  \left[ \frac{Sinh^2(\pi k_B T \tau) Sinh^2(\pi k_B T(t-\tau))}
{(\pi k_B T)^3 Sinh(\pi k_B Tt)} \right]^{-\alpha} \! \! \! \! +h.c. \, . 
\label{FWMT}
\end{eqnarray}
For large $t$, $F_{FWM}(t)$ presents an exponential decay $exp[-\alpha \pi k_B Tt]$
which becomes a power law decay $t^{-\alpha}$ at zero temperature.

\begin{figure}
\psfig{figure=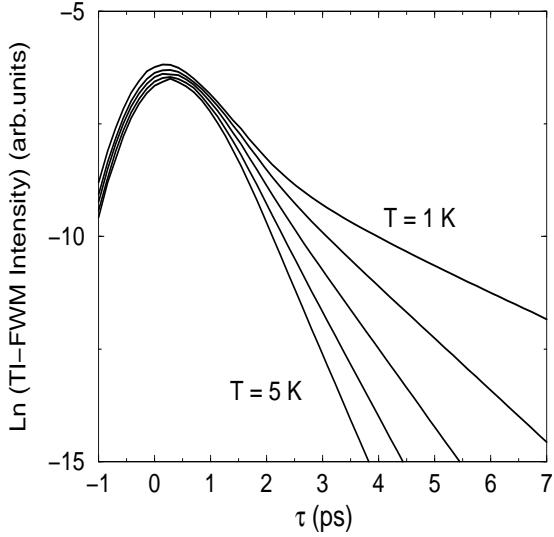,height=8.0cm,width=8.0cm}
\caption{TI-FWM signal as a function of $\tau $ (in ps) for $\alpha =0.7$ 
and $\Delta t=0.7$ps at different temperatures from $1$ up to $5$K 
by successively increasing $T$ in $1$K steps.}
\label{delay}
\end{figure}

Usually, the TI-FWM intensity as a function of $\tau$,
$I_{FWM}(\tau)=\int dt |F_{FWM}(\tau,t)|^2$, is
measured in the experiments \cite{Bar}. In order to obtain realistic results
beyond the delta-like pulses approximation, we have performed numerical
integrations of Eq. (\ref{nlsignal1}) 
with Gaussian pulses having $\Delta t=0.7$ps and $\alpha=0.7$
as shown in Fig. \ref{delay}, for different temperatures.
The maximum is located around $\tau=0$, for which the
overlap of the laser pulses is maximum. 
$I_{FWM}(\tau)$ can show non-exponential
relaxation for $1/\pi k_B T > \tau >> \Delta t$,
in exact analogy to the case of the
dephasing of CC oscillations discussed in section III. For 
$\tau >> \Delta t, 1/\pi k_B T$,
it can be analytically shown from our calculation of
$\chi^{(3)}$ that the decay is exponential, of the form 
$e^{-2 \alpha \pi k_BT \tau}$. 
It must be pointed out that the two different regimes of the TI-FWM 
as a function of $\tau $ shown in Fig.\ref{delay} have been observed 
experimentally \cite{Berkeley} in the presence of a magnetic field.

\section{DISCUSSION and CONCLUSION}

The main concern of this paper is the temporal evolution of the laser
induced optical  coherence of a doped semiconductor in the regime where FES is
observed. In marked contrast with undoped semiconductors, the induced
coherence decays, even at zero temperature, without the intervention
of any inelastic scattering or statistical broadening. We refer to
this decay as intrinsic dephasing. Its origin lies in the excitation
of a continuum of low energy conduction electron-hole pairs whenever
a hole is promoted from the valence band to the conduction band.  In
the spectral domain, these low energy excitations can give rise to
the FES. In the time domain they produce the intrinsic dephasing. 

We have presented calculations of the optical response of a doped
semiconductor, as modeled by Hamiltonian (\ref{MND}), in some
standard experimental situations. The question is whether the
physical processes not included in that Hamiltonian will obscure  our
predictions. There are three additional sources of decay of the
optical coherence which can compete with the 'intrinsic dephasing':
electron-electron (e-e) scattering, electron-phonon  scattering and
inhomogeneous broadening of the localized valence hole levels
\cite{Shahbook}. 

A rough estimate of the decay time of the optical coherence due
to  e-e scattering , 
$T^{e-e}_2$,  can be obtained as
the inverse  of the scattering rate of electrons at $\epsilon_F$. 
For two-dimensional electrons it has been shown \cite{Giuliani} that 
$(T^{e-e}_2)^{-1} \propto T^2 log(T)$, at $k_B T << \epsilon_F$. At
low $T$, this e-e dephasing is less important than the FES intrinsic
dephasing (linear in $T$).  Employing a Thomas-Fermi approximation,
we can estimate $T^{e-e}_2=15$ps at 10 K and $\epsilon_F=20$meV, 
showing a dephasing much slower than $T^{FES}_2 = 1/\alpha \pi k_B T
= 0.35$ps, for $\alpha=0.7$ and the same temperature.  The effect of
electron-electron interaction in the nonlinear optical response of
doped samples has been considered in more detail  by Primozich {\em
et al.} \cite{PerakisPRB,PerakisReport}
for pump-probe experiments, where $|{\cal E}_2|>>|{\cal E}_1|$. 

The electron-phonon interaction will also have a contribution, mainly
due to the scattering between conduction electrons and acoustical
phonons, which are the relevant lattice excitations at low energies.
This interaction can be described by a deformation potential
Hamiltonian \cite{Mahan}, which implies a cubic dependence on
temperature of the scattering rate $(T^{ph}_2)^{-1} \propto T^3$. We
have performed an estimate of this dephasing time, which yields
$(T^{ph}_2)^{-1} = 80$ns for an electron at $\epsilon_F=20$meV,
$T=10$ K, in a GaAs quantum well.

Thus, both electron-electron and electron-phonon effects give rise to
{\em slower} decays of the optical coherence so that they  will not
compete with the FES intrinsic dephasing at low temperatures.

The decay of the optical coherence due to the broadening in the
distribution of  the hole energies depends on the  particular details
of each sample. However, this dephasing is quite  independent on
temperature.  In the case in which inhomogeneous broadening is more
efficient than intrinsic FES dephasing, the experimental study of
$(T_2)^{-1}$ as a function of temperature  would allow to separate
the linear term, $(T_2^{FES})^{-1}$, which is the most important
temperature dependent
contribution, as we have shown.  

Hence, it is our contention that the dynamics of the optical coherence
of a doped sample in the FES regime, as described in this work,
can be observed. However, the limitations of both the Hamiltonian,
the bosonization and the perturbative expansion call for further work
on the theory side.

From the experimental point of view, the realization of the
experiments suggested in this paper would
permit independent measurement of the singularity exponent $\alpha$,
as well as the observation of new physical phenomena, like
phase-shift in the CC oscillations (section III) or the FES in time
integrated FWM signal as a function of the exciting frequency near the
Fermi energy.

In summary, we have presented a theory for the transient optical
response of the FES. The use of the bosonization to describe the low
energy excitations across the Fermi level allows the analytical
evaluation of the linear and nonlinear response both at zero and
finite temperature. CC of the energy absorbed at resonance with the
FES can be performed.  CC oscillations show a phase-shift which
depends on the singularity exponent, $\alpha$. The FWM signal shows a
sharp asymmetric  resonance near $\epsilon_F$ as a function of the
exciting energy, and is strongly suppressed with temperature.     We
have shown that both CC and FWM experiments could be used to study
the decay of the laser induced coherence or dephasing. In contrast to
the case of undoped samples, the bath of Tomonaga bosons responsible
of the FES produces a new dephasing mechanism which depends linearly
on temperature.

Work supported in part by MEC of Spain under contract PB96-0085, Fundacion
Ram\'on Areces and CAM under contract 07N/0026/1998. Diego Porras thanks
Spanish Education Ministry for its FPU grant. J. Fernandez-Rossier
acknowledges Spanish Ministry of Education for its postdoctoral
fellowship.

\appendix

\section{General expression for the average of n polarization operators}

In this appendix we prove Eq. (\ref{chin}), which allows to calculate the
nonlinear optical susceptibilities at any order $n$.
First of all we factorize the correlation function into different bosonic modes:

\begin{eqnarray}
\langle B(t_0) B^{\dagger} (t_1) &&... B(t_{n-1}) B^{\dagger} (t_n) \rangle 
\nonumber \\ = &&
\prod_{k} 
\langle  B_k(t_0) B_k^{\dagger} (t_1) ... B_k(t_{n-1}) B_k^{\dagger} (t_n) \rangle
,
\label{factorink}
\end{eqnarray}

where $B^{\dagger}_k(t) = exp [ \beta_k^{*}(t) b_k^{\dagger}- \beta_k(t) b_k ]$, with
$\beta_k(t) = (1+V \rho) e^{i \frac{k}{\rho} t}$. We ignore for the moment
the index $k$ and define $\beta_j = \beta(t_j)$.
Using the well known relation $e^A
e^B = e^{A+B}e^{\frac{1}{2}[A,B]}$ we can easily show that:

\begin{eqnarray}
\langle B(t_0) && B^{\dagger} (t_1) ... B(t_{n-1}) B^{\dagger} (t_n) \rangle =
 \nonumber \\
  \prod_{j>i=0}^{n}&&e^{-i Im(\beta_i^{*} \beta_j) (-1)^{i+j+1}}
\langle 
     e^{- \sum_{i} \left( (-1)^{i} \beta^{*}_i b^{\dagger}  -h.c. \right) } 
                                                                      \rangle .
\label{factor1}
\end{eqnarray}

The average in (\ref{factor1}) is calculated assuming a thermal distribution of
bosons:

\begin{eqnarray}
\langle 
     e^{- \sum_{i} \left( (-1)^{i} \beta^{*}_i b^{\dagger}  -h.c. \right) }
 \rangle = \hspace{3.5cm} &&  \nonumber \\ 
e^{- (1/2 +  N_B(k))|\sum_{j} \beta_j (-1)^j |^2 } .  
\label{factor2}
\end{eqnarray}

We expand the absolute value inside of the exponential in (\ref{factor2}):

\begin{eqnarray}
|\sum_{j} \beta_j (-1)^j |^2 
 = 2 \sum_{j>i} (|\beta|^2- Re(\beta_i^{*} \beta_j))(-1)^{i+j+1} .
\label{factor3}
\end{eqnarray}

We have $Re(\beta^{*}_i \beta_j)= (1+V \rho)^2 cos \frac{k}{\rho}(t_i-t_j)$ and
$Im(\beta^{*}_i \beta_j)= (1+V \rho)^2 sin \frac{k}{\rho}(t_i-t_j)$.
Substituting (\ref{factor2}) into (\ref{factor1}) and writing explicitly the momentum
index, $k$, we obtain:

\begin{eqnarray}
&&\langle B(t_0) B^{\dagger} (t_1)...B(t_{n-1}) B^{\dagger} (t_n) \rangle  
=  \nonumber \\
&&\prod_{j>i=0}^n \left[ 
  exp \left( - \alpha 
   \sum_{k} ((1+2 N_B(k))(1-cos \frac{k}{\rho}(t_i-t_j)) + \right. \right. \nonumber \\
&& \left. \left. \hspace{3.5cm} 
   i \, \, sin \frac{k}{\rho}(t_i-t_j) ) \right)  \right]^{(-1)^{i+j+1}} 
       = \nonumber \\
&&\prod_{j>i=0}^n \langle B(t_i) B^{\dagger} (t_j) \rangle ^{(-1)^{i+j+1}} .
\label{factor4}
\end{eqnarray}

The factorization formula implies that $\chi^{(n)}$ of Hamiltonian
(\ref{MND}) can be expressed as a product of linear ($\chi^{(1)}$)
susceptibilities, when one restricts to the low energy spectrum
(that is, resonant
excitation at the FES and long time response). A very similar factorization
is found in other physical problems
in which a localized level interacts with the low-energy excitations of an
electron bath, such as the Kondo effect \cite{Yuval}
or an impurity in a Luttinger liquid \cite{Delft}.
In both cases, the factorization formula allows
to write a perturbation expansion in a parameter that plays the role of the
electric field in the FES case.

%%%%%%%%%%%%%%%%%

\widetext


\begin{references}

\bibitem[\dagger]{dagger} Present address: Department of Physics.
University of California San Diego, La Jolla, CA92092 US.

\bibitem{Mahan} G.D. Mahan, {\it Many-Particle Physics}, Plenum, New York 
(1981). 

\bibitem{experiments}
M.S. Skolnick,
J.M. Rorison, K.J.Nash, D.J. Mowbray, P.R. Tapster, S.J. Baas, and A.D. Pitt, 
Phys. Rev. Lett. {\bf 58}, 2130 (1987);
%%
I.V. Kukushkin,
K. von Klitzing, and K. Ploog, 
Phys. Rev. B {\bf 37}, 7788 (1989);
%%
W. Chen,
M. Fritze, W. Walecki, A.V. Nurmikko, D. Ackley, J.M. Hong, and L.L. Chang, 
Phys. Rev. B {\bf45}, 8464 (1992);
%%
K.J. Nash,
M.S. Skolnick, M.K. Saker, and S.J. Bass, 
Phys. Rev. Lett.
{\bf 70}, 3115 (1993);
%%
J. Rubio,
H.P. van der Meulen, J.M. Calleja, R. Bergmann, V. Haerle and F. Scholz, 
Phys. Rev. B {\bf55}, 16390 (1997); 
%%
S.A. Brown,
J.F. Young, Z. Wasilewski, and P.T. Coleridge, 
Phys.  Rev. B {\bf56}, 3937 (1997).

\bibitem{Herko}
H.P. van der Meulen,
I. Santa-Olalla, J. Rubio, J.M. Calleja, K.J. Friedland, R. Hey and K. Ploog,
Phys. Rev. B {\bf 60}, 4897 (1999).

\bibitem{Schotte}K.D. Schotte and U. Schotte, Phys. Rev. {\bf 182}, 479
(1969).

\bibitem{Shahbook}J. Shah, {\it Ultrafast spectroscopy of semiconductors 
and semiconductor nanostructures}, Springer, Berlin (1996).


\bibitem{Kim} D.-S. Kim, J. Shah, J.E. Cunningham, T.C. Damen,
S. Schmitt-Rink and W. Schafer, 
Phys. Rev. Lett. {\bf 68}, 2838 (1992).


\bibitem{Bar} S. Bar-Ad, I. Bar-Joseph, Y. Levinson, and H. Shtrikman,
 Phys. Rev. Lett. {\bf 72}, 776 (1994).

\bibitem{Brener} I. Brener, W.H. Knox and W Sch\~affer, Phys. Rev.
B {\bf51}, 2005 (1995).

\bibitem{Perakis1} I.E. Perakis and D.S. Chemla, Phys. Rev. Lett. {\bf72},
3202 (1994).

\bibitem{PerakisPRB} N. Primozich, T. V. Shahbazyan, and I. E. Perakis, Phys. Rev.
B {\bf 61}, 2041 (2000).

\bibitem{PerakisReport} I.E. Perakis and T.V. Shahbazyan, Surf. Sci. Reports
{\bf 40}, 1 (2000).


\bibitem{CC}   A.P. Heberle, J.J. Baumberg, K. K\"{o}hler, Phys. Rev. Lett.
{\bf 75}, 2598 (1995). 

\bibitem{CC2} X. Marie, P. Le Jeune, T. Amand, M. Brousseau, J. Barrau, and M.
Paillard, Phys. Rev. Lett. {\bf 79}, 3222 (1997).  

\bibitem{CC3} J. Fernandez-Rossier, C. Tejedor and R. Merlin, 
Semicond. Sci. Technol. vol 15, R65-R80 (2000). 


\bibitem{Nozieres} P. Nozi\`eres and C.T. De Dominicis, Phys. Rev. 
{\bf 178}, 1097 (1969).

\bibitem{Uenoyama}
T. Uenoyama and L.J. Sham, Phys. Rev. Lett. {\bf 65}, 1048 (1990).


\bibitem{Ohtaka} K. Ohtaka and Y. Tanabe, Rev. Mod. Phys. {\bf 62}, 929 (1990).

\bibitem{chi3excitons} V.M. Axt and Stahl, Z. Phys. B {\bf 93}, 195
(1994); {\bf 93}, 205 (1994). Th. \"Ostreich, K. Sch\"onhammer and L.J.
Sham, Phys. Rev. B {\bf58}, 12920 (1998).


\bibitem{Bastard} G. Bastard, {\it Wave mechanics applied to semiconductor
heterostructures}, les \'editons the physique, Paris (1988).



\bibitem{Mukamel} S. Mukamel, {\it Principles of nonlinear optical spectroscopy},
Oxford University Press, Oxford (1994). 

\bibitem{Wehner} M.U. Wehner, M.H. Ulm, D.S. Chemla, and M. Wegener,
Phys. Rev. Lett. {\bf 80}, 1992 (1998).

\bibitem{note1} $\tau <0$ means that the pulse labeled as 
$2$ appearing twice in 
Eq. (\ref{nlsignal1}) arrives to the sample before than the pulse $1$ 
appearing only once in Eq. (\ref{nlsignal1}).

\bibitem{Berkeley} N.A. Fromer, C. Sch\"uller, D.S. Chemla, T.V. Shahbazyan, I.E.
Perakis, K. Maranowski, and A.C. Gossard, Phys. Rev. Lett. {\bf 83}, 4646 
(1999).

\bibitem{Giuliani} G.F. Giuliani and J.J. Quinn, Phys. Rev. B {\bf 26},
4421 (1982).

\bibitem{Yuval} P.W. Anderson and G. Yuval, Phys. Rev. Lett {\bf 23}, 89 (1969).

\bibitem{Delft} J. von Delft and H. Schoeller, Annalen der Physik {\bf 4},
225 (1998).

\end{references}
\end{document}